# Enhancement Programming Skills and Transforming Knowledge of Programming through Neuroeducation Approaches

Spyridon Doukakis, Panagiotis Vlamos
Department of Informatics
Ionian University
Greece

*Abstract*— **Programming digital devices and developing software is an important professional qualification, which contributes to employment opportunities. Despite this fact, there is a remarkable shortage in suitable human resources. In this context, research studies focus on issues of programming didactic, teaching models, programming paradigms, which are meant to enhance and optimize programmers' skills. Recent development of brain imaging techniques such as electroencephalography and the functional magnetic resonance imaging, have provided additional opportunity for neuroscientists to explore the functional organization of the human brain. With the use of these techniques, this research is an approach to supporting learning in the field of learning and teaching computer programming. On one hand, there is an attempt to connect theoretical neurosciences with cognitive science; on the other hand, the obtained research data will contribute to the identification of practices that can be applied to formal and informal programming education.**

## I. INTRODUCTION

Over the past several decades, knowing to program digital devices and to develop software has been an important professional qualification, which contributes to employment opportunities. Despite the fact that in recent years there has been a significant increase in the demand for professionals specialized in programming, in well-paid posts, it can be observed that there is a remarkable shortage in suitable human resources [1]. Furthermore, it is observed that students in computer science departments find it difficult to cope with the demands of programming courses, resulting in students' aversion form deepening into programming. One result of this aversion is the partial reinforcement of the labor market with new programmers [2]. Moreover, in the context of education aiming at software development and device programming, both in formal and informal education, it is observed that there is a variety of approaches, which may begin with selecting a specific programming paradigm (e.g. imperative approach) and continue with different paradigms (object-oriented and/or visual programming) or follow other learning pathways [3].

In this context, studies have been carried out by researchers who focus on issues of programming didactic, teaching models, learning theories pertaining to the development of students' programming skills, teaching environments, and programming paradigms, which are meant to enhance and optimize programmers' skills [4] [5] [6]. These studies have contributed to the improvement of programming courses in the context of formal and informal education, methods of teaching textual and visual programming as well as the imperative and object-oriented approach. The evidence obtained are usually related to good teaching practices and interventions, which signify ways of action but have little impact on the courses constructed in various curricula, because some of these research findings are difficult either to generalize or be adopted by others [2].

Concurrently, in recent years, neuroscience attempts to explain the workings of the brain and the related nervous system, the functional architecture of the mind, and how the brain and mind map work together. The field is contributing to our basic understanding of the neural mechanisms underlying human development and learning. Recent development of brain imaging techniques such as electroencephalography (EEG), positron emission tomography (PET) and functional magnetic resonance imaging (fMRI) have provided additional opportunity for neuroscientists to explore the functional organization of the human brain. In light of these advances, neuroscience has experienced rapid growth over the last three decades and tended to form links with other disciplines. Education is one such discipline which, by incorporating neuroscience, can enhance our understanding of mental and physiological processes involved in learning. The long attempts towards connecting neuroscience, cognitive science, psychology, and education have resulted in the emergence of a growing interdisciplinary field of study which has been labeled as neuroeducation. In the field of neuroeducation, educational experts, cognitive scientists and neuroscientists collaborate aiming to implement neuroscientific research findings in educational contexts [7] [8] [9].

In the next section, the educational context and the techniques of teaching computer programming are outlined. The third section connects the neuroscience with education, so as to describe the neuroeducation. The fourth section presents current research and the context of our research. Finally, the paper concludes with the impact that this approach may have to the education and teaching of programming.

## II. Teaching Computer Programming

Computer (more generally digital device) programming is a basic skill in the field of computer science and it is also the passage from theoretical to applied computer science. In addition, computer programming enhances both the computational and algorithmic thinking of trainees during problem solving [10]. Computer programming is mandatory to all computer science curricula. Moreover, due to the increased demand for professionals in computer programming in Europe and the USA, numerous training programs have been developed and made available at no cost for beginners in programming digital devices, as well as for unemployed individuals with little experience in computer programming.

However, most students, particularly in the first year of study, find programming difficult to grasp, have difficulties in decomposing problems, developing plans and implementing their plans with programming languages to solve programming problems, let alone master. At the same time, many adults fear training courses about programming. As a result, many of the first year students and adults that attend computer programming courses fail, which is leading to high dropout rates from the courses [2]. The weaknesses that trainees have concerning programming may be associated with problem solving ability [11] or the absence of viable mental models of key programming concepts which may result in misconceptions and difficulties when learning to program [12]. Moreover, they face difficulties in tracing, reading and understanding pieces of code and learning the syntax and semantics of a language at the same time.

Nevertheless, in all these programs, the way of teaching computer programming is a very important and crucial factor of formal and informal education and in the context of learning and developing trainees' skills. To this end, several teaching approaches have been proposed, which are based, on one hand, on the capabilities and beliefs/opinions of curriculum designers and, on the other hand, on the market needs. As a result, computer programming is not only difficult to learn, but also difficult to teach.

The complexity of teaching programming has emerged in the literature and has been identified as one of the challenges in computer science education [13]. Pears et al. describe four categories relating to course development: curriculum; pedagogy; language choice; and tools for supporting learning [14]. In specific, they identified the lack of an accepted context, of a methodology that would contribute to the design, development, enhancement and implementation of programming courses. Moreover, they identify an abundance of empirical data and small-scale studies that, nevertheless, do not contribute to a possible generalization. Finally, they concluded that the two basic issues that affect design are the choice of programming paradigm and the programming language.

At the same time, education in computer programming is realized using appropriate representations while developing different representations in the trainees. According to research in the area of teaching and learning, when students can interact with a suitable representation, their learning is enhanced [15]. However, special emphasis has been given to the learning and teaching approach with the use of more than one representations [16]. In this context, it is common for trainee programmers to become familiar with two types of computer programming, namely the textual and visual programming, using appropriate programming languages. The goals of education in these two manners are for the students, on one hand, to become fluent with both approaches and, on the other hand, to develop the skill to select the appropriate type depending on the needs of specific application (e.g. mobile applications). In this way, trainees get to know different representation codes and consequently multiple representations that may reinforce their learning. Thus, it is common for programmers to attempt the implementation of the same algorithmic approach or even the same part of the program for different needs and consequently with a different programming type (textual or visual). The textual and visual programming are the two basic programming types that students and adults are trained in, but they are also the two types on which the existing software and its applications available on the market are based upon.

The two types of programming, the learning and teaching environment, the kind of programming, are likely to have a different impact on the trainee, since it is possible to conjecture that these parameters can affect the professional involvement of prospective programmers in programming. Therefore, the acquaintance of trainees with multiple representations, different teaching environments and the ways they are involved in them has led researchers to explore: a) the design of educational programs, b) the ways that these programs serve and support learning, and c) the tasks which the trainee has to engage in when interacting with multiple representations and different programming types [17] [15] [18] [8].

It is evident from the above that there have been several studies related to the ways and methods of teaching programming. Nevertheless, it can be observed that there is a shortage of studies that might bring forth ways of cognitive reinforcement of trainees in programming and learning pathways that favor programming education. This can be achieved through collection of data regarding observed brain activity (in real time) and their correlation with the trainees' cognitive performance. This will enable modeling of the learning process and understanding the trainees' brain activity while working with programming activities, with a view to providing valid and applicable recommendations for educators regarding what and how to teach programming.

## III. NEUROEDUCATION

The cognitive processes taking place during programming are assessed on the general cognitive theories of problem solving, which include processing of structural and semantic information, acquisition of knowledge in fragments, construction of information in schemata and solving problems of design [19]. Moreover, one of the dominant research techniques in the psychology of programming adopts the methodology of trying hypotheses with the use of small groups of individuals who carry out the same tasks in a clearly defined context. The purpose of these studies is usually to measure the duration and precision of execution of one or more tasks. Comparison of groups may provide statistically significant differences, but, to date, has stopped short of delineating different strategies applied by trainees while programming, which can offer considerable data for understanding cognitive functions. Various approaches can be used to tackle the problem, such as: a) carrying out experiential studies and collecting data to be utilized in qualitative analysis, b) conducting longitudinal studies of specifically defined educational contexts in programming or c) long-term observation of trainees as they develop software [17]. At the same time, with a view to investigate the efficiency of different representations, emphasis has been given to the sensory channel and/or the modality of representations (i.e. visual, auditory and tactile) [15].

Moreover, researchers propose as promising methods the imaging techniques that can contribute to linking cerebral activity to the physical and cognitive activity of the participants. More specifically, in a recent study the researchers used fMRI to measure program comprehension [20]. They observed 17 participants inside an fMRI scanner and they found five brain regions, which are related to working memory, attention, and language processing that fit well to the understanding of program comprehension. In another fMRI study the researchers examined code comprehension, code review and prose review [21]. According to the results of their experiment, which was carried out with 29 participants, it was revealed that the neural representations of the programming languages in relation to the natural languages are separate. With the use of EEG, researchers measure programmer expertise and showed that electrical activity in the brain can indicate prior programming experience by class level, and self-reported experience levels [22]. The researchers concluded that with the EEG measures of cognitive load can quantify programming task performance across a spectrum of expertise, and that cognitive demands differ across expertise levels.

In a recent study, eye-trackers were used for the study of the cognitive process during problem solving and the comprehension of the code of programs [18]. The researcher studied two important parameters: the type of representation (textual programming versus visual programming) and the gender of trainees. Results showed that the gender and the type of representations used by programmers influence their productivity and efficiency. Eye-trackers were also utilized while studying the comprehension of Unified Modeling Language (UML) class diagrams [23].

Other interesting studies make use of suitable questionnaires that explore the trainees' satisfaction when working with textual versus visual programming for Arduino applications [5] or when working on designing software [24]. Research findings regarding the trainees' work with the use of multiple representations have yielded diverse results. For example, with specific reference to novice trainees and possibly because of the cognitive load [25], it is stipulated that they need to manage through the short-term/working memory in order to understand the constructions, the semantics, the relationships, while correlating, translating and integrating the new representation in the previously available representations [15] [16], so it seems that multiple representations are an obstacle to their education.

The above interesting and useful findings indicate the significance of studying different mental states through observation of possible changes in brain activity, brain rhythms in particular, with criteria of electrophysiology, as trainees (of formal and informal education) are learning or working as programmers with various types of programming, which are also different representations.

Such studies fall within the boundaries of the interdisciplinary field of neuroeducation, where neuroscientists, cognitive scientists, psychologists and educators cooperate with a view to contributing to the theoretical and practical understanding of learning. Neuroeducation, according to Paul Howard–Jones (in [9], p. 56), "better reflects a field with education at its core, uniquely characterized by its own methods and techniques, and which constructs knowledge based on experiential, social and biological evidence" [26]. Therefore, neuroeducation attempts to contribute to the basic educational research, ultimately aiming to influence the way of teaching in the classroom. Moreover, it tries to contribute conclusions to issues of educational practices that are important for supporting the trainees. Finally, recording brain activity offers the possibility to identify the neurodevelopmental differences that affect educational results and also identify individual differences in the trainees' brain that contribute to reflecting the level of learning according to the curriculum. The development of appropriate techniques for the "imaging" of the brain and of the way it performs in different cognitive functions will contribute to a better understanding of the basic functions that are related to learning and reinforce education. Research shows that understanding cerebral processes during training can contribute to the enhancement and attainment of learning [27].

## IV. RESEARCH CONTEXT

The proposed research study falls within the scope of cognitive neuroscience and neuroeducation. In this research data collection will allow the assessment of cognitive function with criteria of electrophysiology. For this purpose, electroencephalography (EEG) will be used as a system for recording brain activity. Thus, real quantitative data will be collected and then the possible correlation between cognitive performance and observed brain activity will be examined.

The purpose of the study is to identify optimal ways for formal and informal education in programming. The study will entail a random sample of participants among undergraduate students who wish to enhance their knowledge of programming. For this reason, novice undergraduate students (students in the first semester of their studies) will participate in a field study in order to explore potential differences in their brain activity during programming with a visual programming language versus a textual programming language. The students will be asked to develop specific programs in a visual programming language and in a textual programming language. The order of these programs will be determined, while the order of languages in which they will work will be differed between the students. With this structure, some of the students will develop a program firstly with a visual programming language with blocks and then with a textual programming language and some others will develop a program firstly with a textual programming language and then with a visual programming language with blocks. The same approach will follow for the other programs, so that to took place all the possible combinations. The above differentiation was chosen in order to make it possible to compare students' brain activity when working on the same program with a visual programming language with blocks relative to a textual programming language and at the same time it will be possible to assess whether their brain activity is influenced by the order of the languages in which they work.

With the utilization of the laboratory structures and the trainees' participation, data will be collected as the participants work with the appropriate activities (programming tasks). The data will be analyzed and the possibility of correlation between their observed brain activity and their cognitive performance in programming activities will be explored. In this way, learning pathways and practices will be identified, which, according to brain activity data, cognitively reinforce the participants' programming skills.

As previously mention, the EEG imaging method will be used to record the brain signal and measure brain activity using a 10/20 system of the standard position of scalp electrodes for a standard EEG record. The EEG method will be used to collect data to evaluate brain signal frequencies associated with computer problem solving and computer programming. The EEG electrodes will be placed in the C4-P4 scalp position. In this research, the BIOPAC data acquisition unit and MP150 and AcqKnowledge 4.3 Software will be used for data acquisition, analysis, storage, and retrieval. Silver chloride electrodes will be applied following the 10/20 system. The EEG will be recorded at 1000 samples/sec with a resolution of 12 bits/sample. Then the data will digitally be filtered using 1-50 Hz band pass filter.

The objectives of the study are:

- To carry out and record the necessary measurements, which will contribute to the development of appropriate educational models for learning programming.

- To identify learning pathways which facilitate programming education, through the study of the observed brain activity and according to the trainees' profile and educational context.

- To record ways that cognitively reinforce trainees during their education in programming, through the study of the observed brain activity and according to the trainees' profile and context.

- To design targeted educational programs of programming teaching, on the basis of learning pathways, the trainees' profile and educational context.

The present study is a modern approach to supporting learning in the field of programming education. On one hand, there is an attempt to connect neurosciences with cognitive science at a theoretical level; on the other hand, the obtained research data will contribute to the identification of practices that can be applied to formal and informal programming education. The identification of practices and learning pathways will enable the development of targeted programs for the reinforcement of programming knowledge or introduction to programming, which will lead to integrate young programmers in the labor market.

V. IMPACT AND CONCLUSION

The proposed study will contribute to the exploration of the most appropriate programming representation and programming environments in different educational contexts (students and adults). Under the aim of this research is the creation of a generalized model of cognitive competences enhancement, which is developed theoretically, is verified experimentally and is based on criteria of electrophysiology and brain imaging [28]. The results can contribute to education by helping reinforce computer science curricula especially in the field of programming. In this way, trainees will have the opportunity to gain proper experience that offers them knowledge and skills, and this will lead to the development of competent programmers for the labor market. Additionally, the results will contribute to the improvement of continuing education programs for adults, to best support retraining in the areas of computer programming and information technology, areas where there is a high demand for specialists internationally, and where properly trained professionals are not limited by lack of jobs in their countries, since they can find opportunities to work remotely [1]. Interventions based on this research can have a profound effect at a very low cost, by enabling the production of more and better trained computer professionals who can be employed directly or who can work remotely.

At the same time, the results will have a positive impact on society. In specific, educating more people in programming and reinforcement of their programming skills also makes it possible for more people to advance their knowledge and skills, so that they can understand digital technology better and more thoroughly and they can intervene in them programwise, especially when these devices include free/open-code software. In addition, the innovative character of the study and the development of an educational model in programming with the use of criteria of electrophysiology and brain activity recording along with charting brain signals will contribute to the international scientific world. Moreover, due to the existing cooperation of the laboratory with similar research centers abroad, collaborations will be made feasible, with a view to further exploitation of the results. Besides, the results may contribute to enhanced sensitization for direct and effective support teaching of individuals presenting underachievement and learning difficulties, since the recording and assessment of performance in cognitive processes is painless and free of possible dangers. In this context, recording cognitive performance in combination with cerebral activity and other cognitive conditions, while simultaneously informing the participant's portfolio/file, can be a good practice for future research endeavors as well as for people's daily lives.

The association of cognitive competence with performance in programming while simultaneously registering cerebral activity in the light of neuroscience will aid understanding the nature of cognitive competence thus contributing to scientific knowledge with reference to cognitive neuroscience. This process will allow the mapping of cognitive competence and the development of an innovative model of assessment of adults' learning skills based on criteria of electrophysiology and related to programming. Therefore, there will be an enhancement of the existing knowledge in this research field (Cognitive enhancement) as regards mapping, studying and associating cognitive competence with cerebral activity [29]. Lastly, the results can contribute to science at large, as they can become a good practice for similar research studies in other educational and professional environments so as to better understand the function of human brain during the individual's engagement in specific actions and activities.